# Lopsided Spiral Galaxies and a Limit on the Galaxy Accretion Rate

Dennis Zaritsky

UCO/Lick Observatory and Board of Astronomy and Astrophysics,
Univ. of California, Santa Cruz, CA, 95064, E-Mail: dennis@ucolick.org

and

Hans-Walter Rix

Steward Observatory, Univ. of Arizona, Tucson, AZ, 85721, E-Mail: rix@as.arizona.edu

## ABSTRACT

We present a measurement of lopsidedness for the stellar disks of 60 field spiral galaxies in terms of the azimuthal $m = 1$ Fourier amplitude, $A_1$, of the stellar light. We confirm the previous result (Rix and Zaritsky 1995) that $\sim 30\%$ of field spiral galaxies in a magnitude limited sample exhibit significant lopsidedness ($\langle A_1/A_0 \rangle \geq 0.2$) at large radii ($R > 1.5$ disk scalelengths).

We conjecture that this lopsidedness is caused by tidal interactions and calculate an *upper* limit on the accretion rate of small galaxies. We exploit the correlation between lopsidedness and photometric measures of recent star formation (Zaritsky 1995) to obtain two independent estimates of the lifetime of these $m = 1$ distortions. First, we show that lopsided galaxies have an excess of blue luminosity relative to that of symmetric galaxies with the same H I linewidth, which we attribute to a recent star formation episode that was triggered by an interaction between the galaxy and a companion. We use stellar populations models (Bruzual and Charlot 1993) to estimate the time since that interaction. Second, we use the N-body simulation of an infalling satellite by Walker, Mihos, and Hernquist (1996) to estimate how fast tidally induced $m = 1$ distortions are erased through phase mixing. Both approaches indicate that the observations are consistent with a hypothesized tidal interaction that occurred about 1 Gyr ago for galaxies that are lopsided at the 20% level. By combining this lifetime estimate for lopsidedness, the observed frequency of such distortions, and a correction to the survey volume that depends on the increase in luminosity during an interaction, we derive an upper limit on the current companion accretion rate of field spiral galaxies (for companion masses $\sim 10\%$ parent galaxy mass) that lies in the range $0.07 - 0.25$ Gyr$^{-1}$. The principal uncertainty in this limit arises from ambiguities in the interpretation of the correlation between lopsidedness and $M_B$.

## 1. Introduction

The growth of galaxies through mass accretion and mergers is an integral part of galaxy formation. The gas consumption problem for late-type galaxies (Kennicutt,

Tamblyn, & Congdon 1994), the ongoing interaction between the Milky Way and the Sagittarius dwarf (Ibata, Gilmore, & Irwin 1994), and the eventual infall of the Magellanic Clouds onto the Milky Way galaxy (Tremaine 1976) all suggest that mass accretion and infall onto disks occurs at the current time. Placing quantitative limits on the rate of such events will further our understanding of galaxy evolution and provide constraints on cosmological models. Great interest was generated by the claim that the vertical thinness of the stellar disks of most spiral galaxies places a stringent limit on recent accretion and hence argues against high $\Omega_0$ models (Tóth & Ostriker 1992). However, recent N-body simulations suggest that this conclusion is complicated by the absorption of the satellite's orbital energy and angular momentum by the halo rather than by the disk (Navarro, Frenk, & White 1994; Walker, Mihos, & Hernquist 1996; Huang & Carlberg 1996), and so the accretion rate remains unknown.

Infalling and merging components presumably span a wide range of masses. Mergers involving comparably sized progenitors, which are often called "major mergers", produce gross morphological distortions (such as large tidal tails; Toomre & Toomre 1972) and are known to significantly increase the star formation rate in the participants (Larson & Tinsley 1978; Lonsdale, Persson, & Matthews 1984; Kennicutt *et al.* 1987; Sanders *et al.* 1988; Lavery & Henry 1988). Extrapolating from major mergers, it is plausible to assume that the two clearest interaction signatures, a disturbed morphology and an elevated star formation rate, should be present at lower levels in the less dramatic, but presumably more common, "minor mergers". There is no degeneracy between an evolved major merger, where the strength of the signatures has faded with time, and a minor merger, because the former will resemble an elliptical rather than a spiral galaxy (Toomre 1977; Schweizer 1982).

Global low-order ($m = 1$ or 2) asymmetries in spiral galaxy disks, the much lower amplitude analogs of tidal tails or bridges, are candidate signatures of mergers and interactions. Baldwin, Lynden-Bell, & Sancisi (1980) first drew attention to the common phenomenon of lopsided neutral hydrogen distributions in galaxies. Richter & Sancisi (1994) studied a much larger sample of galaxies by connecting lopsidedness in the spatial distribution of neutral hydrogen to asymmetries in the 21 cm line profile. They concluded that at least 50% of the galaxies in H I surveys have significant asymmetries. Rix and Zaritsky (1995; hereafter RZ) found and quantified significant lopsidedness in the *stellar* distribution of spiral galaxies. Lopsidedness is usually not as dramatic in the stars as in the neutral hydrogen, but its presence in the stellar distribution demonstrates that the asymmetry in these cases is not the result of an interaction with the gaseous intergalactic medium (as seen, for example, in NGC 4654; Phookun & Mundy 1995). Furthermore, the stellar asymmetry occurs at smaller radii than the H I counterpart and hence has a shorter dynamical time and lifetime. Consequently, lopsidedness in the stellar distribution may prove to be a finer chronometer of the interaction phenomenon. Finally, if interactions generate a burst of star formation, then the spectral energy distribution of the stellar population provides an external test of the kinematically-derived chronology. The ability to derive two independent timelines for the interaction allows us to cross-check estimates of lifetimes and corresponding accretion rates.

*We will estimate the minor-merger rate by identifying a possible signature of interaction and measuring both its frequency and lifetime.* The interaction signature we choose to use is the $m = 1$ distortion of disk galaxies. A numerical simulation (Walker, Mihos, & Hernquist 1996, §3.6) demonstrates that the infall of a small satellite can produce a large $m = 1$ distortion. Observationally, we need to determine the distribution of $m = 1$ amplitudes for a fair sample of galaxies. Theoretically, we need to (1) calibrate the $m = 1$ amplitude as a function of perturber mass, orbital parameters, and time since the interaction, and (2) determine what other mechanisms, such as an internal disk instability, might contribute to this measure. There is some theoretical work on $m = 1$ distortions and instabilities (Sellwood & Merritt 1994; Weinberg 1994; Syer and Tremaine 1996), but it is yet unclear whether long-lived $m = 1$ modes can arise and be sustained in real disks (with or without the aid of a perturbing companion). *An estimate of the accretion rate by attributing all of the lopsidedness to recent accretion and mergers will lead to an upper limit to the accretion rate.* In this study, we focus on the observational task and peripherally treat the theoretical task.

We measure the distribution of lopsidedness in field spiral galaxies from a moderately sized sample of nearby galaxies (43 new galaxies and 17 from RZ) and develop a simple measurement of asymmetry based on the azimuthal Fourier decomposition of the galaxy light. In §2 we describe the sample, the observations, and the numerical analysis of the data. In §3 we discuss the frequency and amplitudes of asymmetries, the search for nearby perturbing companions, lifetime estimates of the asymmetry using both stellar population synthesis models (Bruzual & Charlot 1993) and a numerical simulation of the infall of a satellite galaxy (from Walker, Mihos, & Hernquist 1996). Finally, we convert these results into an upper limit on the current satellite accretion rate for field spiral galaxies. In §4 we summarize our results.

## 2. The Data

Our sample of nearly face-on target galaxies consists of galaxies of Hubble type S0 or later, at declinations $\delta < +10°$ between $3h < \text{RA} < 16h$, brighter than $m_B = 13.5$, with redshifts $v < 5500$ km s$^{-1}$, and with kinematic inclinations less than $32°$ ($\cos i = 0.85$). The kinematic inclination is determined from the "inverse Tully-Fisher" relation: $\sin i \approx \frac{W_{20}}{2v_c(M_B)}$, where $W_{20}$ is the observed H I linewidth, $v_c(M_B)$ is the mean circular velocity for galaxies of absolute magnitude $M_B$, determined from more edge-on systems, and $M_B$ is calculated using a smooth Hubble flow model with H$_0 = 75$ km

s$^{-1}$ Mpc$^{-1}$. The data necessary to compile such a sample was drawn from the H I catalog of Huchtmeier and Richter (1989) and the RC3 (de Vaucouleurs *et al.* 1991).

Both the I (0.8$\mu$m) and K$'$ (2.2$\mu$m) observations were obtained with the Las Campanas 1m Swope telescope. The I-band images were taken with a 2048×2048 thick Tektronix CCD, which has a spatial scale of 0.69 arcsec pix$^{-1}$, during the nights of 14 and 15 February 1994. The exposure times for all of the I frames were 900 sec and the sky was nearly photometric with only slight occasional cirrus. The K$'$-band images were taken with a 256×256 NICMOS3 IR array; reimaging optics provided a spatial scale of 0.92 arcsec pix$^{-1}$. These data were obtained during four photometric nights, 16 – 19 February 1994. Typical K$'$ exposure times were 1500 sec on object, with an equal amount of time off object. There are some slight variations in exposure times due to scheduling constraints and mechanical or software problems.

The observed galaxies are listed in Table 1. One galaxy in Table 1 (NGC 5006) does not satisfy the previous selection criteria because of slight differences in galaxy classifications between the H I catalog (Huchtmeier & Richter 1989), which was used for the selection, and the RC3 catalog (de Vaucouleurs *et al.* 1991), which was used to compile the Table. In addition, several galaxies have inclinations $> 35°$ listed in Table 1 because their kinematic inclinations differ from their photometric inclinations as given in the H I catalog and Table 1. The inclinations in the Table for NGC 3873 and ESO 317-G020 are estimated from our images using the disk axial ratios, rather than being drawn from the Huchtmeier & Richter compilation because the cataloged inclinations were grossly different from the apparent inclinations.

Both the I and K$'$ data were reduced using IRAF[1]. The I-band images were bias subtracted and flat fielded using twilight sky flats. Photometric calibration was done with Landolt (1992) standards, observed typically at the beginning and end of each night. The K$'$ -band data were reduced as described by RZ. We use the RZ analysis software to measure the disk scalelengths and azimuthal Fourier components of the light distribution at both I and K$'$ . The center is defined to be the brightest point near the center of the image (the galaxies all have well-defined nuclei). This choice can be justified a posteriori by the lack of large measured asymmetries in the central regions (cf. RZ), and by the agreement between the values derived from the I and K$'$ images (see §2.1).

Our choice for a quantitative measure of lopsidedness is the average of the ratio of the $m = 1$ to $m = 0$ Fourier amplitudes, $A_1$, between 1.5 and 2.5 scalelengths. We refer to this quantity as $\langle A_1 \rangle$ and calculate it from both our I and K$'$ galaxy images. The uncertainty in $\langle A_1 \rangle$ is set to be the *rms* variation of the value of $A_1$ within the radial interval. To reduce the effect of foreground stars on the measurement of $\langle A_1 \rangle$, we have replaced all pixel values above a selected threshold (which is set to be roughly equivalent to the central surface brightness of the galaxy) with the sky value. We have also reanalyzed the data presented by RZ (except for NGC 1015 which is a strongly ringed galaxy) in order to calculate $\langle A_1 \rangle$ for those galaxies (see Table 2). Our measurements of $\langle A_1 \rangle$ range from 0.020 to 0.352 and we define galaxies with $\langle A_1 \rangle \geq 0.2$ as significantly lopsided.

---

[1]IRAF is distributed by the National Optical Astronomical Observatories, which are operated by AURA, Inc., under contract to the NSF.

We present $\langle A_1 \rangle$ rather than $\tilde{A}_1$ (the uncertainty corrected value; because $A_1$ is a positive definite quantity, errors bias the measurement upward and a correction is necessary (see RZ for details)) because the correction to $A_1$ depends on the uncertainty in $A_1$, which we do not know precisely for each galaxy. If the uncertainty in $\langle A_1 \rangle$ is $\sim 0.06$ (a conservatively large estimate; see below), then an observed value of $\langle A_1 \rangle$ of 0.2 implies a true value of 0.19. Evidently, this correction is small for the lopsided galaxies of most interest (*i.e.*, $\langle A_1 \rangle \gtrsim 0.2$).

## 2.1. Estimating Uncertainties

We now demonstrate that although the true uncertainties in $\langle A_1 \rangle$ are likely to be larger than the formal uncertainties, they are unlikely to be significantly larger than 0.06. Therefore, as discussed above, the uncertainties will not significantly affect $\langle A_1 \rangle$.

First, to estimate the error in $\langle A_1 \rangle$ due to incompletely masked stars in the I-band images, we reanalyze the images of two symmetric (*i.e.*, small $\langle A_1 \rangle$) galaxies (NGC 2223 and ESO513-G015) to which we have added random stellar fields drawn from the corners of four other images. Except for the three cases (out of 32) in which an extremely bright star landed near the galaxy, the average increase in the measured $\langle A_1 \rangle$ is 0.017 for NGC 2223 and 0.042 for ESO513-G015.

Second, we compare the results for the 18 galaxies with both I and K′ data (Figure 1). The standard deviation of $\langle A_1 \rangle$ measured in I and K′ about a line of unit slope is 0.063, which again suggests that the $\langle A_1 \rangle$ uncertainties are likely to be somewhat larger than those estimated from the variance of $A_1$ between 1.5 and 2.5 scalelengths. However, this estimate of the uncertainty in $\langle A_1 \rangle$ includes wavelength dependent variations in morphology. The best fit line to the data in Figure 1 has slope $= 0.98 \pm 0.03$ if the line is forced through the origin. The agreement of this fit with a line of unit slope indicates that there is no systematic dependence of $A_1$ with wavelength between the I and K′ bands. We concentrate on the I-band data because they extend to fainter surface brightness levels and span a larger sample of galaxies.

Third, we compare the two galaxies (NGC 1703 and IC 2627) that were observed both for this study and by RZ. For NGC 1703, the measurement of $\langle A_1 \rangle$ from the current sample data is $0.096 \pm 0.011$ in the I-band and is $0.105 \pm 0.010$ from the K′ data of RZ. The results for IC 2627 are $0.248 \pm 0.021$ and $0.174 \pm 0.017$ from the current I and K′ band data respectively, and $0.293 \pm 0.025$ from the reanalyzed RZ data. The average of the K' measurements of $\langle A_1 \rangle$ is 0.234±0.015, which agrees with the I-band value, although the difference in the K' measurements suggest that $\langle A_1 \rangle_{K'}$ can have uncertainties $\gtrsim 0.06$. We expect the K' measurements to have larger uncertainties than the I-band measurements because the disk beyond two scalelengths is weakly detected in our K' data.

These various tests demonstrate the basic reliability of the $\langle A_1 \rangle$ measurements, although they suggest that the true uncertainties in $\langle A_1 \rangle$ range from 0.04 to 0.06 (rather than the formal value $\sim 0.02$). We adopt the formal uncertainties for the Tables and Figures, but caution that the true uncertainties may be 2 to 3 times larger than those quoted.

## 3. Results and Discussion

### 3.1. The Distribution of $\langle A_1 \rangle$

The distribution of $\langle A_1 \rangle$ is shown in Figure 2 for both the new and the RZ samples. Despite the differing wavelengths of the observations ($0.8\mu$m for the current sample and $2.2\mu$m for the previous sample), the two distributions are similar and confirm the general correspondence between I and K' measurements of lopsidedness. Because we modified our measure of lopsidedness and removed the ring galaxy NGC 1015 from the RZ sample, there are now fewer highly lopsided galaxies ($\langle A_1 \rangle \geq 0.2$, by the present definition) in the RZ sample than shown in their Figure 3. Even so, three (of 17) objects in their sample have $\langle A_1 \rangle \sim 0.2$. The results from the present sample confirm the RZ result that about 1/3 (14 of 43 or 33%) of "normal" field spirals in a magnitude limited sample are significantly lopsided. In the combined sample, 16 of 60 galaxies, or 27%, are significantly lopsided. If we have no implicit preference for selecting lopsided galaxies, then these numbers represent the characteristics of field spiral galaxies. In §3.3 we discuss why lopsided galaxies may be over-represented by up to a factor of four in our sample. Figure 3 (Plate 1) presents I-band images for some of the more lopsided galaxies in our sample. As can be seen from these images, the lopsidedness is due to extended features rather than to localized cites of recent star formation.

### 3.2. Nearby Companions

If lopsidedness lasts a relatively short time ($\sim 1$ Gyr), as suggested by winding arguments (RZ) and by the numerical simulations discussed below, then the asymmetry is not a relic from the main galaxy formation epoch but is instead due to a later interaction or merger. For example, the stellar disk of M51 is lopsided as a consequence of the current interaction with its companion (Rix and Rieke 1993). None of the galaxies in our sample are clearly interacting, so where are the companions? Either, the companion galaxy has merged with the primary, or it is speeding away from from the primary. By assuming a companion velocity of 150 km s$^{-1}$ relative to the primary (cf. Zaritsky *et al.* 1993), by adopting a mean age of 0.5 Gyr for the distortions, and by averaging over all viewing angles, we estimate that the typical perturber in a hyperbolic encounter has a projected separation of $\sim 7$ arcmin from its primary at the average primary distance (35 Mpc). Therefore, most "escaping" perturbers should lie within our CCD images, which are 20 arcmin on a side.

For the companion galaxy search we used the SExtractor package (Bertin 1995) to identify galaxies with total I-band apparent magnitude $< 17$ within our images. Because redshifts are unavailable for most of the projected neighbors, we depend on the projected separation to estimate their distance from the primary galaxy. We find no significant correlation between the projected density of galaxies with $M_I < -16$ (absolute magnitudes are calculated assuming the projected companions are at the same distance as the primary) and $\langle A_1 \rangle$. However, this lack of a correlation with *projected density* does not exclude a correlation of lopsidedness with *true local density*. As shown by spectroscopic surveys for satellite galaxies (Zaritsky *et al.* 1993), satellite galaxies are a small ($\sim 3\%$) fraction of all projected neighbors at these magnitudes. We identified a total of 231 candidate satellites, which implies that about eight of the candidates are true satellites. Because the counting statistics generate uncertainties that are larger than the unknown eight satellites (15), it is difficult to determine any property related to the true satellites. Furthermore, we selected against obviously interacting systems by selecting galaxies with normal morphological classifications.

One conclusion that we can draw from our images is that significant lopsidedness does not require a close, current interaction (cf. Figure 3). If lopsidedness arises from interactions, then either the lopsidedness lasts a sufficiently long time for perturbers to have drifted away from the primary, or the perturbers have been accreted by the parent galaxy. Models for the generation of lopsidedness must include both accretion and nearby passages. *By assuming that the lopsidedness we observe is due to accretion, rather than nearby passages, we will obtain an upper limit on the disk accretion rate.*

### 3.3. The Star Formation/Lopsidedness Connection

By analogy to the starbursts found in major mergers, one might suspect that there should be enhanced star formation in strongly lopsided galaxies (cf. Zaritsky 1995; hereafter Z95). To examine this hypothesis, we combine the current data with those of RZ to maximize our sample size. As discussed in Z95, a "mass-normalized color" is the most appropriate measure of the star formation activity for this work because it minimizes the dependence of the color measurement on morphology and luminosity. We define the mass normalized color via the Tully-Fisher relationship (Tully & Fisher 1977; hereafter T-F), which predicts the absolute blue magnitude of the galaxy from its neutral hydrogen linewidth. We adopt the B-band T-F relationship given by Pierce and Tully (1992): $M_B = -19.55 - 7.48(\log W_R^i - 2.5)$, where $W_R^i$ is H I linewidth corrected for internal turbulence and inclination as outlined by Tully (1988). The difference between the observed and predicted blue magnitudes (assuming pure Hubble flow and $H_0 = 75$ km s$^{-1}$ Mpc$^{-1}$) is taken to be the "excess" blue light from the galaxy and indicative of the current and recent ($<$ few Gyr) star formation activity. We refer to this quantity as $\Delta$B or the peculiar luminosity. Negative values of $\Delta$B indicate that the galaxy is more luminous than expected for its rotation speed.

To search for a correlation between lopsidedness and recent star formation, we plot $\langle A_1 \rangle$ versus $\Delta$B for the combined sample in Figure 4. We have applied a 0.61 mag offset to the magnitudes from the RC3 (de Vaucouleurs *et al.* 1991) relative to those from the ESO-LV catalog (Lauberts & Valentijn 1989) to place all of the magnitudes on the same system. (This offset is derived from the comparison of the magnitudes for galaxies in the current sample drawn from the two sources). A Spearman rank correlation test indicates that $\langle A_1 \rangle$ and $\Delta$B are correlated with 96% confidence ($R = -0.37$). If, instead, the magnitude offset between the two samples is set so the average value of $\Delta$B is equal in the two samples, then the Spearman test indicates that a correlation is present at the 98% confidence level ($R = -0.42$). The clear outlier at the top of Figure 4, ESO 580-G022, is retained because it has no other particularly distinguishing characteristic (although it is one of the two galaxies in the sample with the latest type, Sdm).

Because several factors may increase the observational scatter, the underlying correlation is probably tighter than it appears in Figure 4: (1) we have combined data from the I and K$'$ band; (2) we have inferred distances from a smooth Hubble flow model; (3) the blue magnitudes differ between RC3 (de Vaucouleurs *et al.* 1991; used for the RZ sample) and the ESO-LV catalog (Lauberts & Valentijn 1989 ; used for the current sample); (4) $\langle A_1 \rangle$ can be underestimated in more inclined galaxies; (5) the inclination corrections to H I linewidths for the nearly face-on sample members (principally galaxies from the RZ sample) are large; and (6) for 14 galaxies there is only a single H I linewidth measurement.

The apparent correlation between $\langle A_1 \rangle$ and $\Delta$B can arise for two reasons: (1) the measurement of $\Delta$B is affected by the asymmetry (*i.e.*, errors in the linewidth or inclination measurement are created by the asymmetry and then produce correlated errors in $\Delta$B ), or (2) the physical mechanism that is responsible for the lopsidedness also affects the stellar populations. As we discuss below, errors in linewidth or inclination appear unable to account for the observed correlation.

Systematic underestimates of the true H I linewidth in asymmetric galaxies will lead to erroneously small (*i.e.*, negative) $\Delta$B's and produce a correlation between $\langle A_1 \rangle$ and $\Delta$B in the observed sense. For the typical coherent radial motions induced by lopsidedness, 7 km s$^{-1}$ (RZ), and the nature of the profiles classified as strongly asymmetric by Richter and Sancisi (1994; their Figure 3), the 30% errors in the linewidth necessary to generate $|\Delta B| = 1$ distortions appear unlikely. Warps in systems with strong lopsidedness would widen the velocity profile of these nearly face-on galaxies and cause $\Delta B$ to increase (opposite to what is needed to account for the correlation).

Systematic overestimates of the inclination will also lead to a correlation in the observed sense. Lopsidedness may, though need not, lead to a nonzero $m = 2$ amplitude in the stellar disk. For illustration, compare a mere displacement of the outer isophote center relative to the nucleus, which introduces no error into the ellipticity measurement but a strong $m = 1$ signal, to the addition of mass, or luminosity, along one axis, which generates a strong $m = 1$ *and* $m = 2$ signal. In the latter case, whether the apparent inclination is increased or decreased relative to the true inclination depends on whether the mass is added along the apparent major or minor axis.

We examine a "worst-case" scenario of the inclination bias by modeling the asymmetry as a stretching of the galaxy along one of the principal directions. We assume that the kinematic inclination is the true disk inclination (*i.e.*, intrinsically $\Delta B = 0$) and that the position angle of the stretching is randomly oriented in azimuth. After averaging over all angles, Figure 5 shows the average induced $\Delta B$ and its dispersion for the sample galaxies.

Only a weak trend in the expected sense is generated (about 0.5 mag over the entire range of $\langle A_1 \rangle$) and the effect on $\Delta B$ is small and nearly symmetric about $\Delta B = 0.25$ for the majority of the sample. The effect is large for a small subset of the sample. In fact, the effect is much larger than the observed spread in $\Delta B$ . The galaxy with the most negative $\Delta B$ (IC 2627) is well modeled by the inclination bias, but otherwise, the generated distribution of $\Delta B$ does not resemble that seen in Figure 4. We conclude that while errors in the inclination may account for a few of the most extreme observed values of $\Delta B$ , they do not account for the observed correlation between $\Delta B$ and $\langle A_1 \rangle$ and, hence, that the correlation between $\langle A_1 \rangle$ and $\Delta B$ principally reflects real variations in the stellar populations of these galaxies.

There are three additional arguments in support of this conclusion. First, the disk ellipticities measured by RZ were small, typically less than 0.1, despite the inclusion of several strongly lopsided galaxies in their sample. This fact, and the large negative values of $\Delta B$ predicted for some galaxies in our sample by our "worst-case" model, indicates that this model for the inclination errors is overly pessimistic. Second, there is a correlation with 99.9% confidence (see Figure 6) between $\langle A_1 \rangle$ and luminosity-normalized color (*i.e.*, $B - R$, another indicator of the effective age of a stellar population). However, because galaxy color and mass correlate (cf. Roberts & Haynes 1995), this result might reflect in part a dependence of $\langle A_1 \rangle$ on Hubble type (although none is seen in Figure 4) or on mass/luminosity (which is seen in Figure 6). To remove the impact of the mass-color relationship, we examine a restricted luminosity range ($-21 < M_R < -19.5$), and find that even for this subsample, the correlation of $\langle A_1 \rangle$ and $B - R$ is detected with 95% confidence. Third, Z95 found a correlation between $\Delta$B and abundance gradients in the sense that galaxies with negative $\Delta$B have steeper gradients. If $\Delta$B decreases systematically with later type or lower mass (as needed to produce the $\langle A_1 \rangle$-($B - R$) relationship), the opposite trend between $\langle A_1 \rangle$ and abundance gradient slope is expected because smaller galaxies tend to have flatter abundance gradients (Vila-Costas & Edmunds 1992; Zaritsky, Kennicutt, & Huchra 1994)).

In summary, the data suggest that $\Delta$B is principally a measure of peculiar luminosity, although other interpretations cannot be entirely eliminated. We proceed by attributing $\Delta$B to stellar population variations and return to this issue after our modeling of the interaction phenomenon.

### 3.4. Lopsidedness, the T-F Relationship, and the Volume Correction

If $\Delta$B variations among galaxies are related to lopsidedness, then there will be systematic biases in measured T-F relationships and morphology statistics of magnitude-limited galaxy samples.

The quoted scatter in T-F studies (*e.g.*, Bernstein *et al.* 1994) is often much less than the $\sim 1$ mag seen in Figure 4. It is unclear to what extent the small scatter in T-F studies is a result of their use of longer wavelength photometry, their smaller inclination errors due to a higher average galaxy inclination, or their selection criteria and reduction procedures, which indirectly eliminate the most lopsided galaxies (for example, Bernstein *et al.* require the ellipticity measurement for each of their galaxies to converge at large radii). This is an important issue, as such culling may not be possible with higher redshift samples.

These difficulties are exacerbated in magnitude limited samples if $\Delta$B and $\langle A_1 \rangle$ are correlated. Lopsided galaxies will be over-represented because they are overluminous. For example, if the significantly lopsided galaxies in our sample are on average brighter by one magnitude than the undisturbed galaxies, then the lopsided galaxies are drawn from a volume that is four times larger than that from which the undisturbed galaxies are drawn. If so, the observed fraction of galaxies in our sample with $\langle A_1 \rangle \geq 0.2$ (0.27) implies that about 7% of field spirals galaxies are significantly lopsided. Such a correction is critical because the inferred accretion rate (assuming a lifetime of 1 Gyr for significant lopsidedness, see below) for a sizable satellite drops from 0.27 Gyr$^{-1}$ to 0.07 Gyr$^{-1}$. The volume correction will remain uncertain until the correlation between $\Delta$B and $\langle A_1 \rangle$ is better understood.

### 3.5. Chronology from Galaxy Colors

The spread in $\Delta B$ suggests that there are stellar populations variations among these galaxies, and the correlation between $\Delta B$ and $\langle A_1 \rangle$ suggests that these variations are related to the morphological disturbance. Using the stellar population synthesis models of Bruzual & Charlot (1993), we calculate the magnitude and duration of color changes for several starburst scenarios.

Our models consist of a star formation burst superposed on an existing population. The initial mass function is a Miller-Scalo (1979) function with upper and lower mass cut-offs of 125 and 0.1 $M_\odot$ respectively (but our broad conclusions are independent of these details). We consider two sets of models. Our "red" models have an old underlying stellar population that formed entirely during the galaxy's first Gyr and a young population that forms during a short, sudden burst 9 Gyr afterwards. Our "blue" models have an underlying population with a constant star formation rate for the last 10 Gyr plus a young population from a sudden, short burst. In both sets of models, the burst forms 10% as many new stars as were present before the burst. This fraction of new stars is meant to mimic the accretion of a satellite with $\sim 15\%$ of the primary's mass and the subsequent episode of increased star formation. We assume that the parent and satellite have similar pre-existing stellar populations and that the parent relaxes to its mass-linewidth relationship instantly after the accretion. We present calculations for three "blue" and "red" models, with burst durations of 0.1, 0.5, and 1 Gyr.

We compile $M_B$ (which is essentially a stellar mass normalized blue color because the model galaxies have the same mass) and $B - R$ as functions of the time since the beginning of the starburst. Results from the "red" models, which show the more extreme color changes, are shown in Figure 7. Some of the models span the observed range in magnitudes (roughly a 3 mag variation in $\Delta B$ , with several of the models spanning the $\sim 2$ mag range seen in the bulk of the galaxy population in Figure 4).

Assuming that $\Delta B = 1$ (a conservative choice for the rightmost edge of the $\Delta B$ distribution shown in Figure 4) corresponds to the plateau in the $M_B$ profiles shown in Figure 7, then we can use population synthesis models to estimate the age since the starburst for galaxies with $\Delta B < 1$. In the upper panel of Figure 8 we show the estimated ages as a function of $\Delta B$ (along a line fitted to the data in Figure 4) based on two of the "red" models: the 0.1 and 0.5 Gyr bursts. We conclude from the top panel of Figure 8 that galaxies with significant lopsidedness probably experienced an interaction well within the last $10^9$ years.

The "blue" models show much smaller color and luminosity evolution and do not reproduce the full observed range of $\Delta B$ . If the "blue" models are appropriate, then $\Delta B$ does not reflect stellar population changes and the volume correction to the frequency of lopsidedness is negligible. The correct model presumably lies between the "red" and "blue" extremes.

### 3.6. Chronology from Merger Simulations

Although the exhaustive theoretical work necessary to establish a "dynamical interaction chronometer" is beyond the scope of this paper, we present one simulation to demonstrate that minor mergers can create asymmetries comparable to those observed in our sample. Walker, Mihos, & Hernquist have generously provided the results from their fully self-consistent stellar dynamical simulation of the accretion of a satellite galaxy which has 10% of the disk mass of the primary galaxy (about 1% of the total mass of the primary — for details see Walker, Mihos, & Hernquist 1996). The satellite begins the simulation at six disk scalelengths from the center of the disk galaxy, in a prograde, circular orbit with an inclination of 30° to the disk plane.

We convert their numerical data to images that are comparable in size and scale to the K′ images. Smoothing and addition of noise is done within IRAF. We take snapshots of their simulated galaxy at times of 1.0, 1.125 and 2.5 Gyr after the start of their simulation (see their Figure 5 for images of the galaxy at the first two of these time steps). Their simulation concludes after 2.5 Gyr. We apply the same analysis algorithm to these images as we apply to our images. We identify the center of the simulated spiral in the last snapshot (the most dynamically relaxed state) and fix the position of the center for the other two images. The values of $\langle A_1 \rangle$ are $0.25 \pm 0.03$, $0.22 \pm 0.02$, and $0.18 \pm 0.01$ for the three times respectively. These results demonstrate that large infalling satellites create $m = 1$ distortions that are entirely consistent with those observed. The lifetime of lopsidedness will depend on many details of the interaction (*e.g.*, mass ratio, disk orientaion, orbital parameters), so the values given above are strictly only valid for the particular case considered. We are currently examining a range of interaction parameter space in order to better estimate the lifetime of lopsidedness.

The simulated measurements of $\langle A_1 \rangle$ as a function of time were used to date galaxies along the line fitted to the data in Figure 4 (cf. bottom panel of Figure 8). The exact relationship between the morphological and stellar population timescales is unclear because we do not know when the star formation rate becomes elevated during an interaction. Nevertheless, the age estimates from the two methods are roughly consistent (offset by $\sim 1$ Gyr at a $\Delta$B of $-2$). If the star formation burst begins a time $T$ after the initial configuration of the N-body simulations (at which time the satellite is at six disk scalelengths from the parent), then the ages in the upper panel of Figure 8 should be increased by $T$. Requiring synchronicity between the kinematic and photometric age estimates sets $T \sim 1$ Gyr (although given the uncertainties this agreement is not required). These dynamical models also suggest that galaxies with significant lopsided have had an interaction within the last Gyr.

We combine the population synthesis models with the results from the dynamical simulation to predict the evolution of recent minor mergers in the $\Delta$B -$\langle A_1 \rangle$ and $(B - R)$-$\langle A_1 \rangle$ planes. Figure 9 shows the evolutionary tracks for both the "red" and "blue" models for a choice of $T = 0.0$ Gyr and a burst duration of 500 Myr. The curves trace the evolution from the beginning of the N-body simulation ($\langle A_1 \rangle$ is extrapolated beyond 2.5 Gyr) and are superposed on the data. In both planes, the models are in gross agreement with or span the existing data. The appropriate models appear to be a combination of our "red" and "blue" extreme models. Figure 9 demonstrates that color and symmetry variations during accretion events should be observable and in the range necessary to explain the data.

### 3.7. Constraining $\Omega_0$

It is well known that the current satellite accretion rate depends on the cosmic density parameter, $\Omega_0$ (*e.g.*, Tóth & Ostriker 1992). From analytic expressions for the number of mergers of certain masses as a function of time (Carlberg 1990; Tóth & Ostriker 1992), we calculate that the probability for the accretion of a lump with $\geq 1\%$ of the primary's mass for a normal spiral galaxy ($\sim 10^{12} M_\odot$; Zaritsky & White 1994) over the last Gyr is 0.46. Therefore, we would expect to see $\langle A_1 \rangle \geq 0.2$ in 46% of the galaxies if (1) the signature of such an accretion is lopsidedness with $\langle A_1 \rangle \gtrsim 0.2$ (as indicated by the simulation), (2) we ignore the very massive interactions that disrupt disks because the number of infalling objects is dominated by those of lower mass, (3) we estimate, based on our two chronometers, that the lifetime of the accretion signature is 1 Gyr, and (4) we postulate that an object accreted by the halo (as calculated by Tóth & Ostriker) is also accreted by the disk. Because the observed frequency of lopsidedness is not that large, the $\Omega = 1$ models and that data can be reconciled by postulating that only a fraction of the halo accretion events become disk accretion events (at most a fraction of 0.15 to 0.59 depending on the adopted volume correction). Such fractions are now not difficult to justify (cf. Navarro, Frenk, & White 1994; Huang & Carlberg 1996). Interestingly, low $\Omega_0$ models may have difficulty reproducing the upper end of our accretion frequency (*i.e.*, assuming no volume correction), especially if only a small fraction of halo mergers become disk mergers.

It is premature to draw inferences on $\Omega_0$ from the present data and interpretations. However, Figure 9 illustrates the potential that observations along these lines have to provide information on the accretion rate, to clarify the connection between star formation and morphological disturbances, and to constrain models of merging and accretion.

The rough agreement between the theoretical merger rates and those derived from our assumption that lopsidedness arises from mergers of satellites that have about a tenth the mass of the parent suggests that our assumption may not be grossly incorrect. Nevertheless, we cannot yet eliminate the possibility that other mechanisms, such as internal disk instabilities, are contributing significantly (or even predominantly) to the observed lopsidedness. Therefore, we stress that the derived merger rates are upper limits.

## 4.    Conclusions

The principal result of this work is the measurement of the $m = 1$ azimuthal distortion in the stellar disks of 60 galaxies and a comparison of the distortions measured at I $(0.8\mu m)$ and K' $(2.2\mu m)$ for 18 galaxies. Our conclusion is that $\langle A_1 \rangle$, the average of the ratio of the amplitudes of the $m = 1$ to $m = 0$ azimuthal Fourier components over the radii from 1.5 to 2.5 disk scalelengths, is greater than 0.2 for about 30% of our magnitude limited sample. $\langle A_1 \rangle$ is evenly distributed between 0 and 0.25.

Based on the examination of our I-band images, we conclude that a close, current interaction is not necessary to produce the large lopsidedness that we observe in some of our galaxies. We conjecture that either the companion has merged with the primary or has receded sufficiently far away from the primary that the interaction is no longer evident.

We compared the measurement of lopsidedness, $\langle A_1 \rangle$, to a measurement of recent star formation activity, the mass-normalized color $\Delta$B . $\Delta$B measures the excess B magnitude of a galaxy relative to the expected B magnitude for a galaxy of that circular velocity (where the luminosity is estimated using a Tully-Fisher relationship). We find that there is a correlation (96% confidence) between $\langle A_1 \rangle$ and $\Delta$B , which suggests that the mechanism by which a galaxy becomes lopsided also affects its luminosity. After examining the possibility that the correlation is due to the effect of lopsidedness on the measures of rotation velocity and inclination, we conclude that the correlation is at least partly due to a real luminosity change that reflects differences in stellar populations among the galaxies.

If lopsided galaxies are systematically brighter than their symmetric counterparts, then they are overrepresented in a magnitude limited sample, and the true fraction of significantly lopsided galaxies may be several times less than the observed $\sim 30\%$. For a 1 mag difference between lopsided and undisturbed galaxies, as might be inferred from our sample (cf. Figure 4), the fraction of lopsided galaxies is lower than that observed by a factor of four. Until the $\Delta$B -$\langle A_1 \rangle$ correlation is understood, large uncertainties in the interaction frequencies will remain.

The correlation between lopsidedness and star formation activity provides two avenues with which to map the timeline of the interaction phenomenon. These lifetime estimates are the critical step in converting the observed frequency of lopsidedness into an estimate of the accretion rate, which can then be used to constrain models of galaxy evolution and cosmology. From population synthesis models (Bruzual & Charlot 1993) with a variety of burst durations and underlying populations, we conclude that galaxies that exhibit $\Delta$B $< -1$ (which corresponds to $\langle A_1 \rangle > 0.15$ along the best-fit line to the data) experienced a 10% burst of star formation (by mass) within the last Gyr. We independently estimate the timescale since the hypothesized accretion of a satellite galaxy using a simulation from Walker *et al.* (1996) and conclude that lopsidedness at the level of $\langle A_1 \rangle \geq 0.2$ (and so a $\Delta B \lesssim -2$) also implies the infall of a 10% satellite less than 1 Gyr ago. Combining our estimate for the lifetime of asymmetries with $\langle A_1 \rangle \geq 0.2$ (1 Gyr) with the observed frequency of such asymmetries in field spirals (between 0.27, if no volume correction is included, and 0.07, if one includes the volume correction for a 1 mag brightening of lopsided galaxies), we conclude that the average infall rate of companions that create such asymmetries in field spirals is at most between 1 every 4 and 1 every 14 Gyr. For a particular choice of the lifetime of lopsidedness, these estimates are upper limits on the accretion rate due to our assumption that lopsidedness arises only from accretion events. The volume correction currently introduces at least a factor of several uncertainty in this limit.

To refine this estimate of the accretion rate, further work on both the observational and theoretical fronts is necessary. Understanding the luminosity-age-$\langle A_1 \rangle$-infalling mass relationships is key in using the observational results to place limits on the disk accretion rate. Volume and absolute magnitude limited samples would remove uncertainties arising from the volume correction and the possible correlation between $\langle A_1 \rangle$ and luminosity. Spectral information, additional colors, or H$\alpha$ imaging, would also help discriminate between a stellar population origin for $\Delta B$ and inclination biases. Further observations involving larger samples will enable us to use the *distribution* of $\langle A_1 \rangle$, rather than simply the number beyond a chosen threshold, as the observational constraint on the accretion rate. Theoretical work is necessary to understand all possible sources of lopsidedness in disk galaxies. The current work suggests that lopsidedness in disk galaxies is a promising technique with which to constrain the accretion rate of galaxies, and thereby constrain the mass evolution of normal galaxies, although significant progress is necessary on several issues before definitive conclusions can be reached.

Acknowledgments: D.Z. acknowledges discussion with Ann Zabludoff about analogous interaction chronometers for E+A galaxies. The authors thank Ian Walker, Chris Mihos, and Lars Hernquist for the use of their simulation. DZ acknowledges partial financial support from the California Space Institute and support provided by NASA through grant number AR-6370.01-95A from the Space Telescope Science Institute, which is operated by AURA, under NASA contract NAS5-26555. We acknowledge the Astrophysical Research Consortium for its support of the NICMOS3 detector used in this work. We thank the Rockwell scientists for their efforts in developing these excellent HgCdTe arrays and the NSF for support. These data were also in part obtained with a charge-coupled device developed by Tektronix for the STIS program for the Hubble Space Telescope.

TABLE 1
GALAXY SAMPLE

| Name | D[a] (Mpc) | T-Type[b] | $i^c$ (°) | $(W_R^i)_T^d$ (km s$^{-1}$) | $(m_B)_T$ | $(m_R)_T^e$ | $\Delta B$ (mag) | $(\langle A_1 \rangle_I)^f$ | $(\langle A_1 \rangle_{K'})^f$ |
|---|---|---|---|---|---|---|---|---|---|
| (1) | (2) | (3) | (4) | (5) | (6) | (7) | (8) | (9) | (10) |
| NGC 1703 | 20.3 | 3.0 | 24 | 141 | 12.31 | 11.43 | −2.32±0.09 | 0.096±0.011 | ... |
| NGC 2196 | 30.9 | 1.0 | 39 | 599 | 11.98 | 10.69 | 1.15±0.10 | 0.043±0.003 | 0.105±0.010 |
| NGC 2216 | ... | 1.7 | ... | ... | 13.70 | 12.34 | ... | 0.115±0.005 | ... |
| NGC 2223 | 36.7 | 3.0 | 28 | 681 | 12.46 | 11.18 | 1.71±0.10 | 0.068±0.006 | 0.142±0.016 |
| NGC 2369B | ... | 3.9 | ... | ... | 14.18 | 12.93 | ... | 0.099±0.011 | ... |
| ESO496-G022 | 31.7 | 7.0 | ... | ... | 13.90 | 12.93 | ... | 0.239±0.034 | ... |
| ESO126-G002 | 39.2 | 3.3 | ... | ... | 13.60 | 12.47 | ... | 0.125±0.014 | ... |
| ESO126-G003 | 38.1 | 4.0 | ... | ... | 13.38 | 12.21 | ... | 0.138±0.013 | ... |
| NGC 2835 | 11.8 | 5.0 | 47 | 242 | 11.03 | 10.32 | −0.66±0.11 | 0.117±0.010 | ... |
| NGC 3084 | 32.5 | 1.7 | ... | ... | ... | ... | ... | 0.293±0.010 | ... |
| ESO317-G020 | 33.3 | 5.3 | 8 | 88 | 13.17 | 12.71 | −1.16±0.10 | 0.248±0.014 | 0.221±0.031 |
| NGC 3241 | 37.8 | 0.0 | ... | ... | ... | ... | ... | 0.073±0.009 | 0.061±0.009 |
| NGC 3244 | 35.2 | 6.0 | ... | ... | 13.09 | 11.83 | ... | 0.106±0.013 | 0.217±0.026 |
| NGC 3250E | 37.6 | 6.0 | 51 | 219 | 13.19 | 12.25 | −1.33±0.10 | 0.196±0.014 | ... |
| NGC 3278 | ... | 5.0 | ... | ... | 13.02 | 11.78 | ... | 0.189±0.019 | 0.155±0.021 |
| ESO317-G054 | 40.7 | 6.1 | ... | ... | 14.01 | 13.88 | ... | 0.289±0.024 | ... |
| NGC 3513 | 15.9 | 5.0 | 34 | 158 | 12.16 | 11.11 | −1.55±0.09 | 0.223±0.008 | ... |
| IC 2627 | 27.8 | 4.0 | 24 | 96 | 12.67 | 11.36 | −3.88±0.12 | 0.248±0.021 | 0.174±0.017 |
| NGC 3783 | 42.7 | 1.5 | 18 | 205 | 12.46 | 11.33 | −0.46±0.10 | 0.042±0.003 | 0.040±0.005 |
| UGCA 247 | 16.7 | ... | ... | ... | 12.98 | 12.00 | ... | 0.306±0.015 | ... |
| ESO320-G024 | 39.8 | ... | ... | ... | 14.12 | 13.24 | ... | 0.157±0.033 | ... |
| NGC 4027 | 22.3 | 8.0 | 37 | 293 | 11.71 | 10.60 | −0.72±0.11 | 0.190±0.027 | 0.244±0.020 |
| ESO267-G016 | 39.8 | ... | ... | ... | 14.21 | 13.18 | ... | 0.220±0.032 | ... |
| ESO506-G029 | 39.7 | 6.0 | 35 | 180 | 14.19 | 13.27 | −1.09±0.09 | 0.243±0.020 | ... |
| NGC 4603D | 37.3 | 6.7 | ... | ... | 14.14 | 12.70 | ... | 0.045±0.006 | ... |
| ESO323-G010 | 39.5 | 4.3 | ... | ... | 13.47 | 12.26 | ... | 0.133±0.007 | 0.163±0.020 |
| NGC 4806 | 32.4 | 4.6 | ... | ... | 13.44 | 12.52 | ... | 0.191±0.015 | ... |
| UGCA 324 | 29.6 | ... | ... | ... | 13.13 | 12.31 | ... | 0.352±0.030 | 0.291±0.040 |
| NGC 4965 | 30.2 | 6.8 | 28 | 467 | 12.76 | 11.96 | 1.18±0.11 | 0.179±0.011 | ... |
| NGC 5006 | 37.0 | −0.7 | ... | ... | 13.33 | 11.89 | ... | 0.054±0.008 | 0.052±0.007 |
| NGC 5156 | 39.7 | 3.5 | 24 | 510 | 12.54 | 11.32 | 0.65±0.11 | 0.104±0.014 | 0.206±0.075 |
| ESO271-G010 | ... | 6.2 | 33 | ... | ... | ... | ... | 0.220±0.012 | ... |
| UGCA 378 | 34.6 | ... | ... | ... | 13.11 | 12.42 | ... | 0.061±0.007 | ... |
| NGC 5489 | 39.6 | 0.5 | ... | ... | ... | ... | ... | 0.093±0.009 | 0.116±0.013 |
| NGC 5494 | 35.9 | 5.0 | ... | ... | 12.64 | 11.36 | ... | 0.168±0.014 | 0.125±0.022 |
| ESO446-G031 | 35.4 | 6.0 | 24 | 237 | 13.65 | 12.63 | −0.47±0.09 | 0.210±0.021 | ... |
| NGC 5556 | 18.4 | 7.0 | 30 | 312 | 12.43 | 11.50 | 0.61±0.10 | 0.142±0.025 | ... |
| IC 4444 | 26.1 | 4.0 | 30 | 305 | 12.25 | 11.08 | −0.40±0.10 | 0.207±0.029 | 0.110±0.015 |
| NGC 5643 | 16.0 | 5.0 | 28 | 370 | 11.03 | 9.92 | −0.07±0.10 | 0.132±0.073 | ... |
| ESO580-G022 | 29.5 | 8.0 | 39 | 243 | 13.80 | 13.01 | 0.16±0.10 | 0.339±0.014 | ... |
| NGC 5757 | 35.1 | 3.0 | ... | ... | 12.77 | 11.52 | ... | 0.083±0.009 | 0.193±0.032 |
| ESO513-G015 | 35.0 | 0.1 | ... | ... | 14.33 | 12.72 | ... | 0.020±0.003 | ... |
| IC 4538 | 30.8 | 5.3 | ... | ... | 12.83 | 11.62 | ... | 0.191±0.019 | 0.150±0.025 |

[a]Distances from pure Hubble flow model with $H_0 = 75$ km sec$^{-1}$ Mpc$^{-1}$. Velocities taken from Huchtmeier and Richter (1989) with the exception of those for NGC 3084, 3241, and 5489 which were taken from Z-cat (Huchra 1993).

[b]T-Types from RC3 (de Vaucouleurs *et al.* 1991).

[c]Inclinations from Huchtmeier and Richter (1989) and references therein.

[d]Velocity widths from Huchtmeier and Richter (1989) and references therein corrected for inclination and internal velocity dispersion as described by Tully (1988).

[e]Total apparent magnitudes from the ESO-LV catalog (Lauberts & Valentijn 1989).

[f]The formal uncertainties listed here are probably underestimates of the true uncertainties (see §2.1)

TABLE 2
GALAXY SAMPLE FROM PAPER I

| Name | $D^a$ (Mpc) | T-Type$^b$ | $i^c$ (°) | $(W_R^i)^d$ (km s$^{-1}$) | $(m_B)_T^e$ | $\Delta B$ (mag) | $(\langle A_1 \rangle_{K'})^f$ |
|------|------|------|------|------|------|------|------|
| (1) | (2) | (3) | (4) | (5) | (6) | (7) | (8) |
| NGC 600 | 24.6 | 6.5 | 30 | 139 | 12.9 | $-1.57 \pm 0.09$ | 0.084±0.016 |
| NGC 991 | 20.5 | 5.0 | 23 | 180 | 12.4 | $-0.83 \pm 0.09$ | 0.224±0.010 |
| NGC 1302 | 22.7 | 0.0 | 18 | 286 | 11.6 | $-0.35 \pm 0.09$ | 0.025±0.003 |
| NGC 1309 | 28.5 | 4.0 | 21 | 360 | 12.0 | $0.31 \pm 0.10$ | 0.144±0.010 |
| NGC 1325A | 17.8 | ... | 21 | 117 | 13.3 | $-1.00 \pm 0.12$ | 0.115±0.019 |
| NGC 1376 | 55.5 | 6.0 | 21 | 415 | 12.8 | $0.12 \pm 0.10$ | 0.163±0.017 |
| NGC 1642 | 61.6 | 5.0 | 28 | 251 | 13.3 | $-1.24 \pm 0.09$ | 0.090±0.005 |
| NGC 1703 | 20.3 | 3.0 | 24 | 140 | 11.9 | $-2.12 \pm 0.09$ | 0.105±0.010 |
| NGC 2466 | 71.5 | 5.3 | 27 | 319 | 13.5 | $-0.58 \pm 0.10$ | 0.070±0.010 |
| NGC 2485 | 61.5 | 1.0 | ... | ... | 13.1 | ... | 0.035±0.004 |
| NGC 2718 | 51.2 | 2.0 | ... | ... | 12.7 | ... | 0.049±0.004 |
| NGC 6814 | 20.9 | 4.0 | 19 | 274 | 12.1 | $0.20 \pm 0.09$ | 0.071±0.015 |
| NGC 7156 | 53.1 | 6.0 | 24 | 249 | 13.1 | $-1.14 \pm 0.09$ | 0.198±0.022 |
| NGC 7309 | 53.3 | 5.0 | 17 | 386 | 13.0 | $0.17 \pm 0.09$ | 0.093±0.011 |
| NGC 7742 | 22.0 | 3.0 | 16 | 264 | 12.4 | $0.26 \pm 0.09$ | 0.044±0.008 |
| IC 2627 | 27.8 | 4.0 | 24 | 96 | 12.6 | $-3.35 \pm 0.12$ | 0.293±0.025 |
| ESO436-G029 | 54.4 | 5.0 | ... | ... | 13.4 | ... | 0.131±0.022 |

$^a$Distances from pure Hubble flow model with $H_0 = 75$ km sec$^{-1}$ Mpc$^{-1}$. Velocities taken from Huchtmeier and Richter (1989) with the exception of those for NGC 3084, 3241, and 5489 which were taken from Z-cat (Huchra 1993).

$^b$T-Types from RC3 (de Vaucouleurs *et al.* 1991).

$^c$Inclinations from Huchtmeier and Richter (1989) and references therein.

$^d$Velocity widths from Huchtmeier and Richter (1989) and references therein corrected for inclination and internal velocity dispersion as described by Tully (1988).

$^e$Total apparent magnitudes from the ESO-LV catalog (Lauberts & Valentijn 1989).

$^f$The formal uncertainties listed here are probably underestimates of the true uncertainties (see §2.1).

## Figure Captions

Figure 1. A comparison of the $\langle A_1 \rangle$ measurements from I and K$'$ images. As suggested in the text, the uncertainty estimates from the standard deviation of $A_1$ between 1.5 and 2.5 scalelengths appears to underestimate the true uncertainty. Nevertheless, there is a good correspondence between the measurements at the two wavelengths and no systematic difference (the slope of the best fit line is 0.98±0.03 if it is constrained to pass through the origin).

Figure 2. The distribution of $\langle A_1 \rangle$ for the current sample is presented in the upper panel (from I-band measurements) and that for the RZ sample is in the lower panel (from K$'$ measurements).

Figure 3 (Plate 1). Two panels of I-band images of six galaxies with $\langle A_1 \rangle \geq 0.2$ from the current sample. The upper panel presents the galaxies at a contrast level surface brightness level to what might be seen on a sky survey photographic plate and the lower panel shows the low surface brightness morphology of the galaxies. The galaxies in the upper row from left to right are NGC 3084, IC 2627, and ESO 267-G016 and in the lower row they are UGCA 324, IC 4444, and ESO 580-G022. In the lower panel, the white dots represent the nuclei of the galaxies. Each image is nearly 6 arcmin on a side.

Figure 4. An examination of the correlation between $\langle A_1 \rangle$ and mass-normalized color, $\Delta$B . In the lower left panel we plot $\Delta$B vs. $\langle A_1 \rangle$ for the combined sample of galaxies (our new sample and the RZ sample). Note that the measurements of the new sample are based on I-band images and those of the RZ sample are based on K$'$ images. A 0.6 magnitude adjustment was made to the RC3 blue magnitudes for the RZ sample based on a comparison of the ESO-LV and RC3 blue magnitudes for the new sample of galaxies. The two sets of points connected by lines represent the galaxies in common between the current study and RZ. In the upper panel we plot the distribution of T-types (morphological type, with late type galaxies having larger corresponding values of the T-type) as a function of $\Delta$B . In the lower right panel we plot $\langle A_1 \rangle$ as a function of T-type.

Figure 5. The effect of inclination errors due to $\langle A_1 \rangle$. The axes are the same as in the main panel of Figure 4. The figure shows the mean and the dispersion (represented by the error bars) of the inferred $\Delta$B that could arise for each galaxy solely due to an incorrectly inferred inclination, as calculated for the "worst case" scenario (see Section 3.3).

Figure 6. The correlation between $\langle A_1 \rangle$ and other photometric properties. In the lower left panel we plot $\langle A_1 \rangle$ vs. $B - R$ for the present sample ($B - R$ colors are unavailable

for the RZ sample). In the upper panel we plot the distribution of $M_R$ as a function of $B - R$ and in the lower right panel the distribution of $\langle A_1 \rangle$ versus $M_R$.

Figure 7. Changes in color and $M_B$ predicted from the "red" population synthesis models. Three "red" models are plotted: a 1.0 Gyr burst model (filled circles), a 0.5 Gyr burst model (filled squares), and a 0.1 Gyr burst model (filled triangles).

Figure 8. Estimated chronology based on $\Delta$B from the Bruzual and Charlot (1993) stellar population synthesis models (upper panel) and on $\langle A_1 \rangle$ from the Walker *et al.* (1996) simulation of the infall of a 10% mass satellite onto a disk galaxy. In the upper panel the ages above the line (in units of Gyr) come from the 0.5 Gyr burst "red" model and the ages below the line come from the 0.1 Gyr burst "red" model.

Figure 9. The evolutionary tracks in the $\Delta$B -$\langle A_1 \rangle$ and $(B - R)$-$\langle A_1 \rangle$ planes are plotted for the 0.5 Gyr burst "red" (solid lines) and "blue" (dashed lines) models. The starburst is assumed to begin when the satellite is six scalelengths away from the galaxy nucleus (the beginning of the N-body simulation, see text for details). The data are superposed for comparison.